# An Inter-Laboratory Comparison of NVNA Measurements

Martin Salter[#1], Laurence Stant[*2], Koen Buisman[§3] Troels Nielsen[$4]

[#]National Physical Laboratory, Teddington, UK
[*]University of Surrey, Guildford, UK
[§]Chalmers University of Technology, Gothenberg, Sweden
[$]Keysight Laboratories, Keysight Technologies, Aalborg, Denmark
[1]martin.salter@npl.co.uk, [2]l.stant@surrey.ac.uk, [3]buisman@chalmers.se, [4]troels_nielsen@keysight.com

*Abstract*—A comparison of nonlinear vector network analyser (NVNA) measurements has been carried out involving four organisations (National Physical Laboratory, UK, University of Surrey, UK, Chalmers University of Technology, Sweden and Keysight Technologies, Denmark). Three nonlinear devices consisting of two amplifiers and a nonlinear verification device (NLVD) were measured by each of the organisations. Results are presented which show generally good agreement between the measurements and give some indication of the typical amount of variability to be expected in measurements of this type.

## I. Introduction

Nonlinear devices, such as power amplifiers, mixers and frequency multipliers, play an important part in modern communications systems. When excited by an incident wave of frequency $f_0$ and of sufficiently large amplitude, a nonlinear device produces scattered waves at frequencies which are harmonics of the frequency $f_0$; the nonlinear device also exhibits compression i.e. the ratios between the complex amplitudes of the incident wave and the scattered waves depend on the power of the incident wave. A nonlinear vector network analyser (NVNA) [1, 2, 3] can be used to test such nonlinear devices. An NVNA measures the amplitudes and phases of the travelling waves at its ports at both the fundamental frequency and at the harmonic frequencies.

A comparison of NVNA measurements has been carried out involving four organisations (National Physical Laboratory, UK, University of Surrey, UK, Chalmers University of Technology, Sweden and Keysight Technologies, Denmark). The purpose of the comparison was to estimate the typical amount of variability to be expected in measurements of this type and also to help the participants to assess the quality of their measurements.

## II. Devices under test, stimulus conditions and measurands

The three devices under test (DUTs) in the measurement comparison consisted of two amplifiers and a nonlinear verification device (NLVD) all of which were fitted with 3.5 mm coaxial connectors. The DUTs were as follows:

- **DUT 1**: A Marki Microwave amplifier [5] with a frequency range of 5 MHz to 26.5 GHz and a nominal gain of 12 dB with 3.5 mm coaxial adaptors on the input and output
- **DUT 2**: The same model of amplifier as for DUT 1 but with 3.5 mm coaxial 6 dB attenuators (matching pads) on the input and output
- **DUT 3**: A non-linear verification device (NLVD) [4]

The stimulus conditions that were applied to the DUTs are listed in Tables I and II. The measurands (quantities to be measured) for DUTs 1 and 2 were: the amplitude and phase of the incident and scattered waves (a and b) at the input and output for five harmonics (including the fundamental), the DC bias voltage and current at the drain and gate, the gain and the power added efficiency (PAE). The measurands for DUT 3 were: the amplitude and phase of the incident and scattered waves (a and b) at the input and output for five harmonics (including the fundamental) and the DC Voltage at the monitor port. The amplifiers were measured under both linear and nonlinear operating conditions depending on the incident RF power level. The nonlinear verification device (NLVD) [4] was designed to be insensitive to the impedance match conditions provided by the NVNA. It consists of three blocks in cascade: an input block comprising a band pass filter to pass the 2 GHz fundamental frequency, a nonlinear block consisting of diodes arranged in a limiting configuration to generate harmonics and an output block to isolate the nonlinear block from the NVNA load port.

TABLE I.  STIMULUS CONDITIONS FOR DUTs 1 AND 2

| |
|---|
| **Fundamental RF measurement frequencies:** 2.4, 3.5 and 5 GHz |
| **Input RF power levels:** − 10 to + 14 dBm in 1 dB steps for DUT 1 and − 4 to + 20 dBm in 1 dB steps for DUT 2 |
| **DC bias conditions:** drain voltage: +7.0 V, gate voltage: -0.3 V |

TABLE II.  STIMULUS CONDITIONS FOR DUT 3

| |
|---|
| **Fundamental RF measurement frequencies:** 2 to 2.4 GHz in 0.1 GHz steps |
| **Input RF power levels:** + 10 dBm |
| **DC bias conditions:** No applied DC bias |

The work described in this paper was funded through the European Metrology Programme for Innovation and Research (EMPIR) Project 14IND10 'Metrology for 5G Communications'. This project has received funding from the EMPIR programme co-financed by the Participating States and from the European Union's Horizon 2020 research and innovation programme.

Some results obtained in the measurement comparison are shown in Figs 1 – 7. In these plots 'lab 1' to 'lab 4' denote the four participants in the comparison. The power and phase for the harmonics of the scattered wave at port 2 ($b_2$) in response to a 5 GHz incident wave at port 1 ($a_1$) are plotted against incident RF power for DUT 2 in Figs 1-3. Similarly, the gain and PAE for DUT 2 at 5 GHz are plotted against incident power in Figs 4-5. The RF incident powers at which the measurements were performed differed slightly for each participant and so the results were corrected by linear interpolation to make them coincide with the requested incident powers. Figs 6 and 7 show the measured power and phase of the scattered wave at port 2 ($b_2$) for DUT 3 plotted against harmonic number for a 2 GHz input signal with a power of 10 dBm.

To compare the variability in the measurements, a standard deviation was calculated for each of the measured parameters. For the complex-valued scattered wave at port 2, the standard deviation of the real and imaginary parts of the complex amplitudes, $S_{re}$ and $S_{im}$ respectively, were combined to give a total standard deviation $s = \sqrt{s_{re}^2 + s_{im}^2}$. The standard deviations obtained for the three DUTs are given in Table III (these values are a maximum taken over all incident powers and all harmonics). The standard deviations given for DUTs 1 and 2 correspond to a 5 GHz input signal whilst that given for DUT 3 corresponds to a 2 GHz input signal.

Lab 2 incorrectly measured the phase of the harmonics of the $b_2$ wave for DUT 1 (apart from the fundamental). This is demonstrated by the reduced standard deviation for $b_2$ when the results of Lab 2 for DUT 1 are excluded. However, excluding the results of Lab 2 for DUT 1 does not affect the standard deviations of gain and PAE because these only depend on the fundamental which was correctly measured by Lab 2. The measurement comparison also brought to light a problem with the measurement by Lab 3 of the amplitude and phase of the harmonics (excluding the fundamental) for DUTs 1 and 2 which was subsequently corrected.

TABLE III.   COMPARISON OF STANDARD DEVIATIONS FOR THE THREE DUTs

| DUT | Scattered wave $b_2$ | Gain | PAE |
|---|---|---|---|
| 1 | 0.06<br>0.012 (excluding lab 2) | 0.4 | 0.9 |
| 2 | 0.0045 | 0.16 | 0.2 |
| 3 | 0.0001 | - | - |

### III. CONCLUSION

The results presented show generally good agreement between the measurements. The variability in the measurements appears to depend on the extent to which the DUT is isolated from the NVNA impedance match conditions with DUT 1 showing the most variability and DUT 3 the least. Some discrepancies in the results highlighted problems with some of the participant's measurements.

ACKNOWLEDGMENT

The authors thank Mamad Rajabi and Dominique Schreurs (KU Leuven) and David Humphreys (NPL) for developing the NLVD and for allowing its use in this measurement comparison.

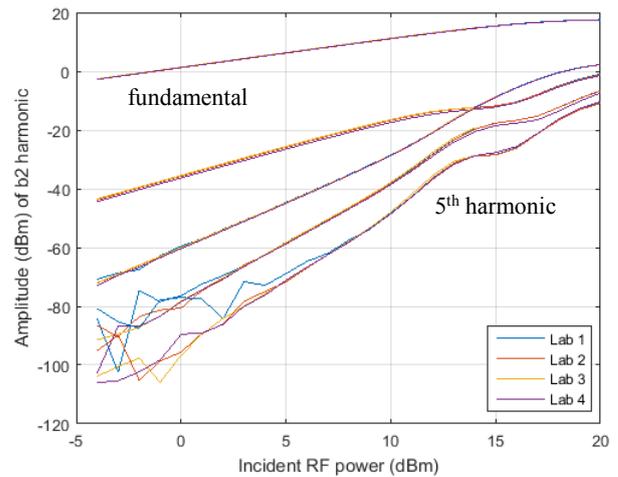

Fig. 1.  Amplitude (dBm) of scattered wave at port 2 of DUT 2 corresponding to incident wave at port 1 of frequency 5 GHz (five harmonics shown)

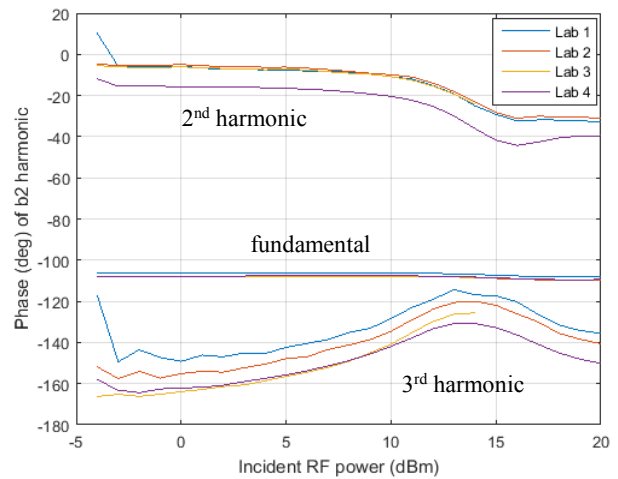

Fig. 2.  Phase (degrees) of scattered wave at port 2 of DUT 2 corresponding to incident wave at port 1 of frequency 5 GHz for harmonics 1 to 3

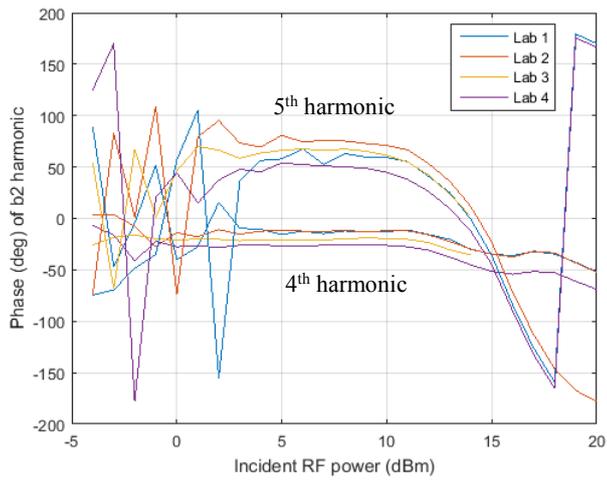

Fig. 3. Phase (degrees) of scattered wave at port 2 of DUT 2 corresponding to incident wave at port 1 of frequency 5 GHz for harmonics 4 and 5

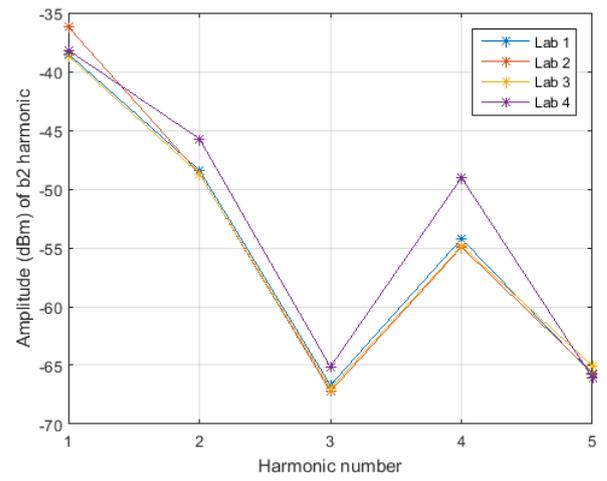

Fig. 6. Magnitude (dBm) of scattered wave at port 2 of DUT 3 corresponding to incident wave at port 1 of power 10 dBm and frequency 2 GHz

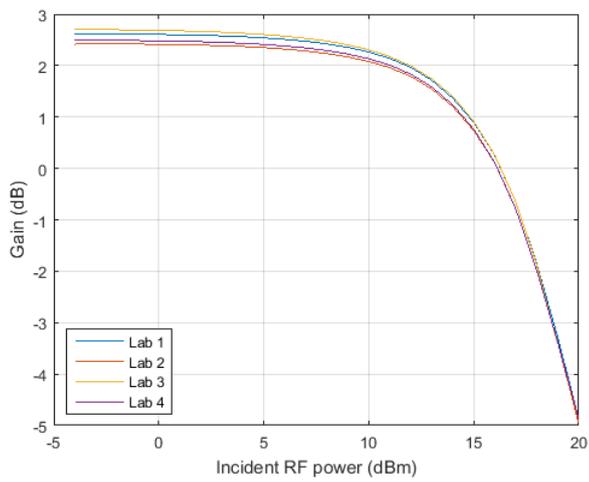

Fig. 4. Gain (dB) of DUT 2 at 5 GHz

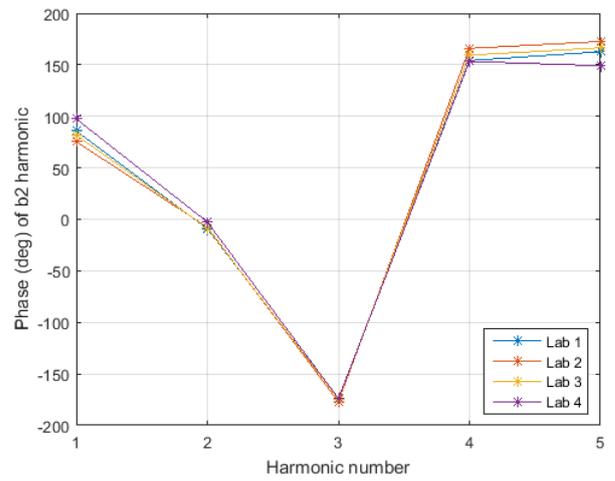

Fig. 7. Phase (degrees) of scattered wave at port 2 of DUT 3 corresponding to incident wave at port 1 of power 10 dBm and frequency 2 GHz

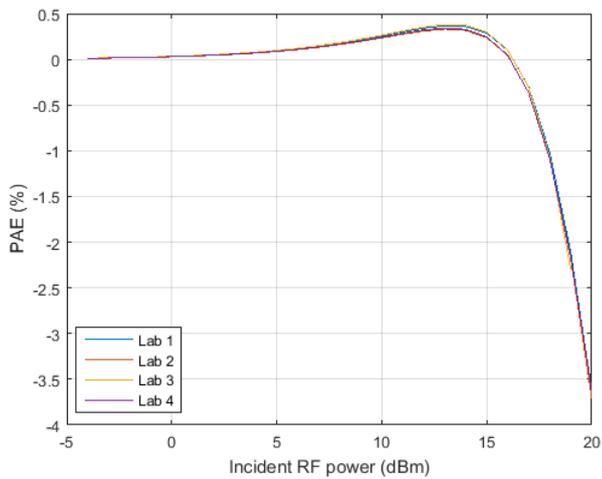

Fig. 5. PAE of DUT 2 at 5 GHz